%% file: draft_30_Dor_C_revised_180710.tex
\documentclass{aastex62}
\usepackage{natbib, aas_macros}
\usepackage{mediabb} 
\usepackage{bm}
\accepted{July 27, 2018}
\submitjournal{ApJ}

\shorttitle{30 Dor C}
\shortauthors{Babazaki et al.}

\begin{document}

\title{A Spatially Resolved Study of X-ray Properties in Superbubble 30 Dor C with $XMM-Newton$}

\correspondingauthor{Yasunori Babazaki, Ikuyuki Mitsuishi}
\email{y\underline{ }babazaki@u.phys.nagoya-u.ac.jp, mitsuisi@u.phys.nagoya-u.ac.jp}

\author[0000-0002-0786-7307]{Yasunori Babazaki}
\affil{Department of Physics, Nagoya University, Furo-cho, Chikusa-ku, Nagoya 464-8601, Japan}

\author{Ikuyuki Mitsuishi}
\affiliation{Department of Physics, Nagoya University, Furo-cho, Chikusa-ku, Nagoya 464-8601, Japan}

\author{Hironori Matsumoto}
\affiliation{Department of Earth and Space Science, Osaka University, Osaka 560-0043, Japan}

\author{Hidetoshi Sano}
\affiliation{Department of Physics, Nagoya University, Furo-cho, Chikusa-ku, Nagoya 464-8601, Japan}

\author{Yumiko Yamane}
\affiliation{Department of Physics, Nagoya University, Furo-cho, Chikusa-ku, Nagoya 464-8601, Japan}

\author{Satoshi Yoshiike}
\affiliation{Department of Physics, Nagoya University, Furo-cho, Chikusa-ku, Nagoya 464-8601, Japan}

\author{Yasuo Fukui}
\affiliation{Department of Physics, Nagoya University, Furo-cho, Chikusa-ku, Nagoya 464-8601, Japan}



\begin{abstract}
We carry out spatially resolved spectral analysis with a physical scale of $\sim$10 pc in X-ray for the superbubble 30 Dor C, 
which has the largest diameter of $\sim$80 pc and the brightest non-thermal emission in superbubbles for the first time. 
We aim at investigating spatial variation of the physical properties of non-thermal emission as detected in some supernova remnants in order to study particle acceleration in a superbubble. 
We demonstrated that non-thermal components are detected in all the regions covering the entire field of 30 Dor C.
The spectra in the west region of 30 Dor C can be described with a combination of the thermal and non-thermal components 
while the spectra in the east region can be fitted with the non-thermal component alone. 
The photon index and absorption corrected intensity in 2--10 keV of the non-thermal component show spatial variation from $\sim$2.0 to $\sim$3.7 and (4--130) $\times$ 10$^{-8}$ erg~s$^{-1}$~cm$^{-2}$~str$^{-1}$, respectively, and the negative correlation between the non-thermal physical properties is observed.
The temperature and normalization of the thermal component also vary within a range of $\sim$0.2--0.3 keV and $\sim$0.2--7 $\times$ 10$^{17}$ cm$^{-5}$ str$^{-1}$, respectively, and the positive correlation between the photon index and the normalization is also detected.
We revealed the correlations in a supperbubble for the first time as is the case in SNRs, which suggests the possibility that the same acceleration mechanism works also in the supperbubble.
\end{abstract}



\section{Introduction} \label{sec:intro}
X-ray and TeV gamma-ray observations clarify that supernova remnants (SNRs) are the acceleration sites of the cosmic rays up to the TeV range 
\citep[e.g.,][]{1995Natur.378..255K,2013Sci...339..807A}.
The accelerated electrons with such energies emit non-thermal X-ray emission via synchrotron radiation which is characterized by power-law energy distribution in an X-ray band. 
Thus, understanding of the non-thermal properties in X-ray is of great importance to study the nature of the accelerated electrons.

The diffusive shock acceleration mechanism \citep[e.g.,][]{2008ApJ...678..939Z} is believed to be a relevant mechanism, 
which can explain the power-law spectrum of the non-thermal X-ray emission. 
Recently, it is reported that several SNRs exhibit the spatial shape variations of the non-thermal X-ray spectra with a physical scale of $\sim$1-5 pc \citep[e.g.,][]{2015ApJ...799..175S,2017ApJ...835...34T}. 
Several authors \citep[e.g.,][]{2015ApJ...799..175S,2017ApJ...835...34T} suggest that the origin of the variations is due to the spatial difference of the cosmic-ray acceleration efficiency related to the surrounding interstellar gas distribution. 

Superbubbles (SBs) are formed by combined phenomena of stellar winds from massive stars in OB associations and eventual supernovae (SNe) of those stars \citep[e.g.,][]{1980ApJ...238L..27B}. 
The morphology of hot gas in SBs is expected to be similar to that of a bubble blown by stellar winds of an isolated massive star \citep{1977ApJ...218..377W}. 
The kinetic energy in some of SBs exceed that in a supernova ($E_{\rm K} \sim10^{51}$ {\rm erg}). 
SBs are filled with hot gas ($\sim$10$^6$ K) heated by stellar wind and SN ejecta.
Non-thermal X-ray emission has been detected from a number of Galactic and extragalactic SBs, 
such as RCW38 \citep{2002ApJ...580L.161W}, Westerlund 1 \citep{2006ApJ...650..203M}, IC 131 \citep{2009ApJ...707.1361T}, N11 \citep{2009ApJ...699..911M}, N51D \citep{2004ApJ...605..751C}, 
and 30 Dor C \citep{2004ApJ...602..257B,2009PASJ...61S.175Y,2015A&A...573A..73K}, 
which suggests a potential accelerating power exceeding an SNR, even though the detection of the non-therrmal emission in N11 and N51D is now doubtful due to the fluctuation of background point sources \citep{2010ApJ...715..412Y}.
However, very few studies have been performed to investigate the spatial variation of the non-thermal X-ray emission in the SBs and thus the cosmic-ray acceleration mechanism in the SBs has yet to be elucidated fully. 

The SB, 30 Dor C, in Large Magellanic Cloud (LMC) was discovered by \citet{1968MNRAS.139..461L}. 
It is believed that 30 Dor C has formed via stellar winds of the LH90 OB association \citep{1993A&A...280..426T} and several SNe. 
The SB has the strongest non-thermal X-ray emission among SBs and a large diameter of $\sim$80 pc.
Thus, it is one of ideal laboratories for studying the non-thermal emission mechanisms associated with a SB. 
The SB includes not only the non-thermal emission but also the thermal emission from the shock-heated interstellar medium \citep[e.g.,][]{2004ApJ...602..257B,2009PASJ...61S.175Y,2015A&A...573A..73K}. 
Therefore, comparing the spatial distribution of the non-thermal and thermal emissions can provide us with hints to reveal association between the acceleration efficiency and the environment potentially.
In this paper, we aim at investigating spatial variation of the physical properties in 30 Dor C with an unprecedented high resolution of $\sim$10 pc. 

The paper is organized as follows: section 2 presents the observations of 30 Dor C with $XMM-newton$ and the data reduction, and sections 3 and 4 describe our analysis method and the results, and the discussions on the spatial variation of the physical properties, respectively. 
In section 5, we summarize our results and discussions. 
At 30 Dor C assuming a distance of $\sim$50~kpc to the LMC, $1\arcmin$ corresponds to 15~pc. 
In this paper, we used HEAsoft v6.21, XSPEC version 12.9, the $XMM-Newton$ Source Analysis Software (XMM-SAS) packaged in SAS 15.0.0 for our spectral analysis and a metal-abundance table tabulated in  \citet{1989GeCoA..53..197A}. 
Unless otherwise stated, the error ranges show the 90\% confidence level from the center value.
\section{OBSERVATION AND DATA REDUCTION}
We, first, retrieved all of the data available for 30 Dor C in the $XMM-Newton$ Science Archive, and then selected the data taken from the pn instrument in European Photon Image Camera \citep[EPIC,][]{2001A&A...365L...1J} with rich ($>$50 ks) net exposure time after the removal of the background flare periods to take advantage of the larger effective area than those of the EPIC-MOS instrument and avoid the systematic error between the detectors. 
\setcounter{table}{0}
\begin{deluxetable*}{cccrc}[b!]
\tablecaption{Observation Log for 30 Dor C \label{tab:obs_list}}
\tablecolumns{5}
\tablenum{1}
\tablewidth{0pt}
\tablehead{
\colhead{Obs. ID} &
\colhead{R.A.} &
\colhead{Dec.} & \colhead{Date} & \colhead{Exposure (ks)\tablenotemark{a}} \\
\colhead{} & \colhead{(J2000.0)} &
\colhead{(J2000.0)} & \colhead{} & \colhead{pn}
}
\startdata
 0104660101 & 05h35m27.99s & -69d16m11.0s & 2000 Nov. 17 & 3  \\
 0406840301 & 05h35m27.99s & -69d16m11.1s & 2007 Jan. 1  & 63   \\
 0506220101 & 05h35m28.30s & -69d16m13.0s & 2008 Jan. 11 & 68  \\
 0556350101 & 05h35m28.30s & -69d16m13.0s & 2009 Jan. 30 & 63  \\
 0601200101 & 05h35m28.30s & -69d16m13.0s & 2009 Dec. 11 & 71 \\
 0650420101 & 05h35m28.30s & -69d16m13.0s & 2010 Dec. 12 & 51 \\
 0671080101 & 05h35m28.30s & -69d16m13.0s & 2011 Dec. 02 & 61  \\
 0690510101 & 05h35m28.30s & -69d16m13.0s & 2012 Dec. 11 & 60  \\
\enddata
\tablenotetext{a}{All exposure times show flare-filtered exposure times.\unskip}
\end{deluxetable*}

The basic information of the observations is shown in Table \ref{tab:obs_list}.
We generated calibrated event files with the SAS tools {\tt epchain}. 
Time intervals with high background rates ($>$ 0.4 cts) seen in light curves of an off-source region in 10--12 keV were discarded in each observation.
The event lists were then filtered further, keeping only 0–-4 patterns in an energy range of 0.4--12 keV.

\begin{figure}[!t]
\vspace{0.5cm}
  \begin{center}
\hspace*{-1.0cm}
    \includegraphics[width=100mm,angle=0]{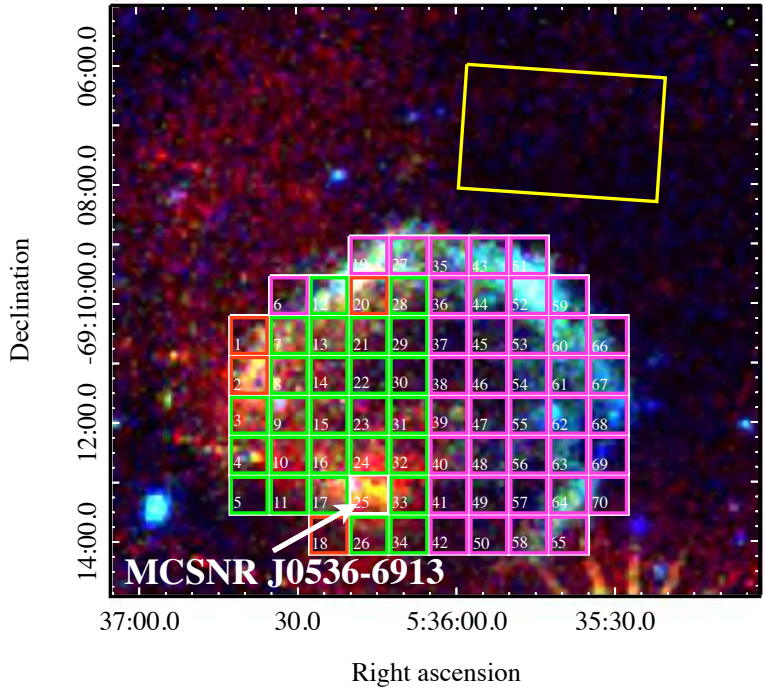}
  \end{center}
  \caption{
$XMM-Newton$ EPIC-pn image of 30 Dor C in 0.3--1 keV (red), 1--2 keV (green),
 and 2--7 keV (blue), respectively. Spectra were extracted from the square regions ($\sim$0.$^\prime$7 $\times$ 0.$^\prime$7) with a region number. 
 One- (non-thermal), two- (one-temperature and non-thermal), and three-component (two-temperature and non-thermal) models are finally adopted in magenta-, green-, and red-color box regions. The background spectrum was extracted from the yellow rectangle. 
}
  \label{fig:30_Dor_C_img}
\end{figure}

\section{Analysis and Results}
In order to conduct spatially detailed X-ray spectral analysis for 30 Dor C, spectrum for each region is extracted from all the seven data and the extracted spectra are fitted simultaneously to reduce the statistical error.
Each spectrum is rebinned to have at least 25 counts per an energy bin to allow the use of the $\chi^2$-statistic.
The energy range in 0.5--7 keV was used in our analysis to avoid detector noise and the EPIC-pn fluorescence line forest just above 7 keV. 
The SAS tasks \texttt{rmfgen} and {\tt arfgen} were utilized to create redistributed matrix files (RMF) 
and ancillary response files (ARF) respectively. 

\subsection{X-ray Background Evaluation\label{analysis_and_results}}
\begin{deluxetable*}{lcc}[b!]
\tablecaption{Best fit parameters obtained by using all seven observations for the background region \label{tab:bgd_para}}
\tablecolumns{3}
\tablenum{2}
\tablewidth{0pt}
\tablehead{
\colhead{Component} &
\colhead{Parameter} &
\colhead{Best-fit Value} 
}
\startdata
\multicolumn{3}{c}{Absorption} \\
\hline 
Galactic (phabs) & $N_{\rm H, Gal.}$ & 0.06 (fixed)\tablenotemark{a}\\
LMC (vphabs)\tablenotemark{b}  & $N_{\rm H, LMC}$  &   0.20${\pm0.16}$       \\
\hline
\multicolumn{3}{c}{Astrophyical X-ray Forground and Background} \\  [2pt]
\hline
LHB (apec)\tablenotemark{c}       & $kT$ (keV)                 & 0.1 (fixed)\tablenotemark{d}\\
               & $Norm$\tablenotemark{e}    & $<$12        \\
GH (apec)\tablenotemark{c}       & $kT$ (keV)                 & 0.22${\pm 0.01}$\\
               & $Norm$\tablenotemark{e}   &  42.5$^{+4.0}_{-4.8}$       \\
CXB (powerlaw)  & $\Gamma$                   & 1.4 (fixed)\tablenotemark{f}\\
               & 2-10 keV intensity\tablenotemark{g}   & 6.0$\pm1.3$ \\
\hline
\multicolumn{3}{c}{Thermal emission in LMC} \\  [2pt]
\hline
ISM (vapec)\tablenotemark{b}   & $kT$ (keV)                  & 0.89$^{+0.04}_{-0.03}$   \\
               & $Norm$\tablenotemark{e}               & 83$^{+17}_{-15}$    \\ 
\hline
$\chi^2/d.o.f$ &                             &  509/375         \\
\enddata
\tablenotetext{a}{Fixed to the Galactic column density from the HI maps \citep[][]{1990ARAA..28..215D}. The unit is 10$^{22}$ cm$^{-2}$.\unskip}
\tablenotetext{b}{Fixed to the representative LMC values \citep[][]{1992ApJ...384..508R,2014PASJ...66...26S}.}
\tablenotetext{c}{Fixed to a solar abundance table tabulated in \citet{1989GeCoA..53..197A}.}
\tablenotetext{d}{Fixed to the value derived from \cite{2009PASJ...61..805Y}. }
\tablenotetext{e}{Normalization of the apec model divided by a solid angle $\Omega$. 
$Norm = (1/\Omega)$ $n_{\rm e}n_{\rm H}$ d$V/[4((1 + z) D_A)^2 ]$ in unit of $10^{14}\ {\rm cm}^5~{\rm str}^{-1}$, 
where, $z$, $n_{\rm e}$, $n_{\rm H}$, $D_A$, and $V$ are the redshift, the electron and hydrogen number densities (cm$^{-3}$), 
the angular diameter distance (cm) and the emission volume (cm$^3$), respectively.}
\tablenotetext{f}{Fixed to an averaged value derived in \cite{2002PASJ...54..327K}.}
\tablenotetext{g}{The unit is 10$^{-8}$ erg s$^{-1}$ cm$^{-2}$ str$^{-1}$.}
\end{deluxetable*}

\begin{figure}
  \begin{center}
\hspace*{-1.0cm}
    \includegraphics[width=100mm,angle=0]{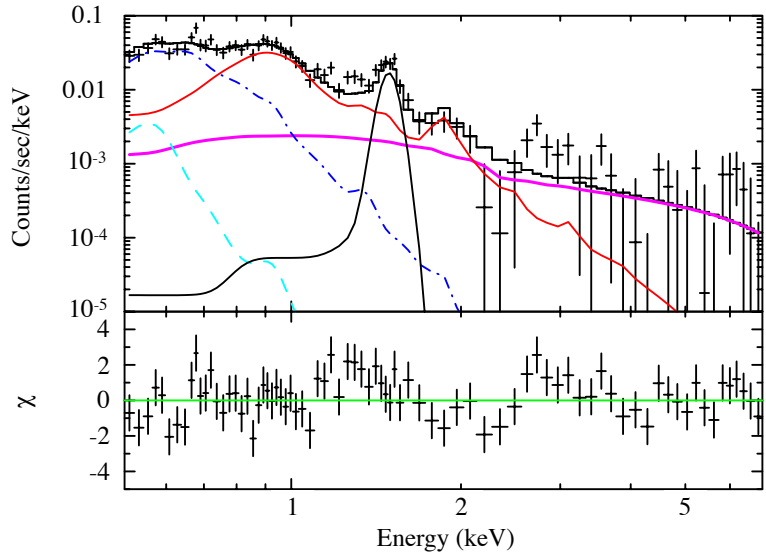}
  \end{center}
  \caption{
A representative spectrum of a background region (Obs. ID: 0601200101) with the best-fit model. The dashed (cyan), dashed-dotted (blue), bold (magenta), solid (red) and solid (black) lines show the LHB, GH, CXB, ISM in the LMC and an artificial Al line, respectively.
 }
  \label{fig:bgd_spec}
\end{figure}
Firstly, we selected a source-free area in the vicinity of 30 Dor C to reduce spatial variation of the detector noise in the field of view shown in figure \ref{fig:30_Dor_C_img} as a background region.
Then, we conducted spectral analysis to confirm whether the region is suitable or not as a background region.
In order to create quiescent particle background (QPB) spectra, we used $XMM-Newton$ Extended Source Analysis Software (XMM-ESAS), packaged in SAS 15.0.0. 
The QPB spectra were subtracted from the spectrum of each region in each observation. \par 
For our spectral analysis, we used the following phyically motivated model:
\begin{eqnarray*}
(apec_{\rm LHB} + phabs_{\rm Galaxy}*(apec_{\rm GH} + \\
vphabs_{\rm LMC} * (powerlaw_{\rm CXB}))
\end{eqnarray*}
The X-ray background emission is comprised of three components \citep[e.g.,][]{2009PASJ...61..805Y}, 
such as an unabsorbed thermal (k$T$ $\sim$0.1 keV) emission 
from the Local Hot Bubble (LHB), an absorbed thermal (k$T$ $\sim$0.2--0.3 keV) emission from the Galactic halo (GH), and an absorbed powerlaw 
\citep[$\Gamma = 1.4$, see ][]{2002PASJ...54..327K} 
which is known as cosmic X-ray background (CXB). 
We used collisionally-ionized optically-thin thermal plasma model APEC \citep{2001ApJ...556L..91S} for the LHB and GH in XSPEC. The metal abundance of these models is fixed to a solar abundance. 
Because the temperature of the LHB component was not constrained well, the temperature is fixed to be a typical value of 0.1 keV \citep{2009PASJ...61..805Y}.
The absorption by our Galaxy and the LMC was also taken into account.
We used a photo-electric absorption model in XSPEC, namely phabs, as the Galactic absorption model. 
The column density $N_{\rm H}$ was fixed at 6 $\times$ 10$^{20}$ cm$^{-2}$ \citep{1990ARAA..28..215D} 
in the direction of 30 Dor C, assuming the solar abundance. 
The absorption by the LMC is modeled with vphabs, in which we can set each metal abundance separately.
The metal abundance was fixed to the representative LMC values \citep[C=0.30 $Z_\odot$, O=0.26 $Z_\odot$, Ne=0.33 $Z_\odot$,][]{1992ApJ...384..508R,2014PASJ...66...26S}, while the absorption column density is set to be free.
The background spectrum, however, can not be described with the model above and there is a significant residual feature around $\sim$1 keV corresponding to emission lines from complex Fe L.
We hence added another thermal component, $apec_{\rm LMC}$, with a different temperature as follows:
\begin{eqnarray*}
(apec_{\rm LHB} + phabs_{\rm Galaxy}*(apec_{\rm GH} + \\
vphabs_{\rm LMC} * (apec_{\rm LMC} + powerlaw_{\rm CXB})).
\end{eqnarray*}
The fitting results improved significantly and the spectrum with the best-fit model is shown in figure \ref{fig:bgd_spec}. 
The best-fit parameters are summarized in Table \ref{tab:bgd_para}. 
The plasma temperature of the added thermal component is consistent with that of the ISM in the LMC \citep[e.g.,][]{2002A&A...392..103S}. 
The 2--10 keV surface brightness of the power-law component 
was $(6.0\pm1.3) \times 10^{-8}$ erg s$^{-1}$ cm$^{-2}$ str$^{-1}$ and the value is in good agreement with the expected CXB intensity \citep{2002PASJ...54..327K}. 
We confirmed that the best fit parameters are consistent with those obtained in each observation and thus all the spectra were fitted simultaneously to reduce the statistical error.
Any further components such as a soft proton model are not required.
Thus, we concluded that the region and model are appropriate to evaluate the X-ray background components including the ISM of the LMC in our analysis.
\begin{figure*}[ht]
  \begin{center}
\hspace{-0.5cm}
    \includegraphics[width=175mm,angle=0]{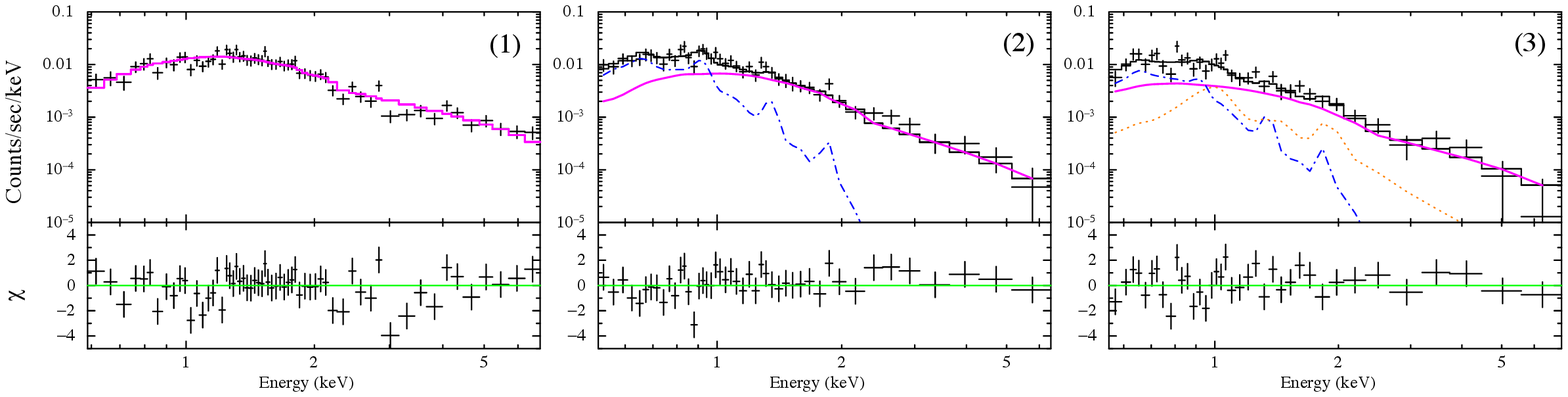}
  \end{center}
  \caption{
Examples of the spectra with the best-fit non-thermal-, two-, and three-, component models. (1) non-thermal-component model for the spectrum in region 52. (2) two-component model for the spectrum in region 13. (3) three-component model for the spectrum in region 1. The magenta solid, blue dashed-dotted and orange dotted lines show non-thermal, low-temperature and high-temperature components, respectively. 
The spectra are extracted from obsid 0601200101.
}
  \label{fig:typical_spectra}
\end{figure*} 

\subsection{Spectral Analysis for the 30 Dor C Field}
In order to investigate spatial variation of the non-thermal X-ray emission in 30 Dor C, 
we divided the 30 Dor C region into 70 regions of 10 pc $\times$ 10 pc (0.$^\prime$7 $\times$ 0.$^\prime$7) grids in unprecedented detail.
The region number is shown in figure \ref{fig:30_Dor_C_img}.
In our spectral analysis, the background spectrum defined in $\S\ref{analysis_and_results}$ was subtracted from each region in each observation.

As indicated in previous studies \cite[e.g.,][]{2004ApJ...602..257B,2009PASJ...61S.175Y,2015A&A...573A..73K}, not only non-thermal emission but also thermal emission is sometimes required at the same time to explain the observed spectra.
Actually, some spectra show significant enhancement around 0.6 and/or 1 keV corresponding to emission lines of highly-ionized oxygen / Fe L-shell complex, respectively.
We attempted to apply three models in the following order, (1) non-thermal model, (2) two-component (non-thermal and one-temperature thermal) model, and (3) three-component (non-thermal and two-temperature thermal) model.
For regions where the fit significantly ($\geq$ 99\% in an F test) improved by adding an additional thermal component, we adopted the two- or three-component model.
A collisionally-ionized optically-thin thermal plasma model, APEC, was used also for the thermal plasma in the regions except a young SNR, MCSNR~J0536-6913, associated with 30 Dor (see the region number 25 in figure \ref{fig:30_Dor_C_img}).
The metal abundance of the low- / high-temperature plasmas mainly emitting oxygen / Fe L-shell lines is fixed to those reported in \citet{2009PASJ...61S.175Y} / \citet{1992ApJ...384..508R}.
The intrinsic absorption column density in the LMC is applied for the both plasmas and linked to that of the non-thermal model.
The only spectrum around MCSNR~J0536-6913 was well expressed with a combination of the non-thermal and non-equilibrium ionization collisional plasma models due to a heavy contamination from the SNR as shown in \cite{2015A&A...573A..73K} and thus we removed the results in our discussion.

Most of the spectra in the east region can be well described with the two- or three-component model, 
whereas most of the spectra in the west region can be well fitted with the non-thermal-component model as shown in Figure \ref{fig:30_Dor_C_img}. 
Figure \ref{fig:typical_spectra} shows examples of the spectra with the best-fit non-thermal-, two-, and three-component models.
We investigated the photon index and intensity of the non-thermal component, the temperature and intensity of the thermal component, and the intrinsic absorption column density of the LMC.
The best-fit parameters in the best fit model are summarized in Table \ref{tab:fitting_results}. 

\begin{figure*}[!t]
  \begin{center}
    \includegraphics[width=175mm,angle=0]{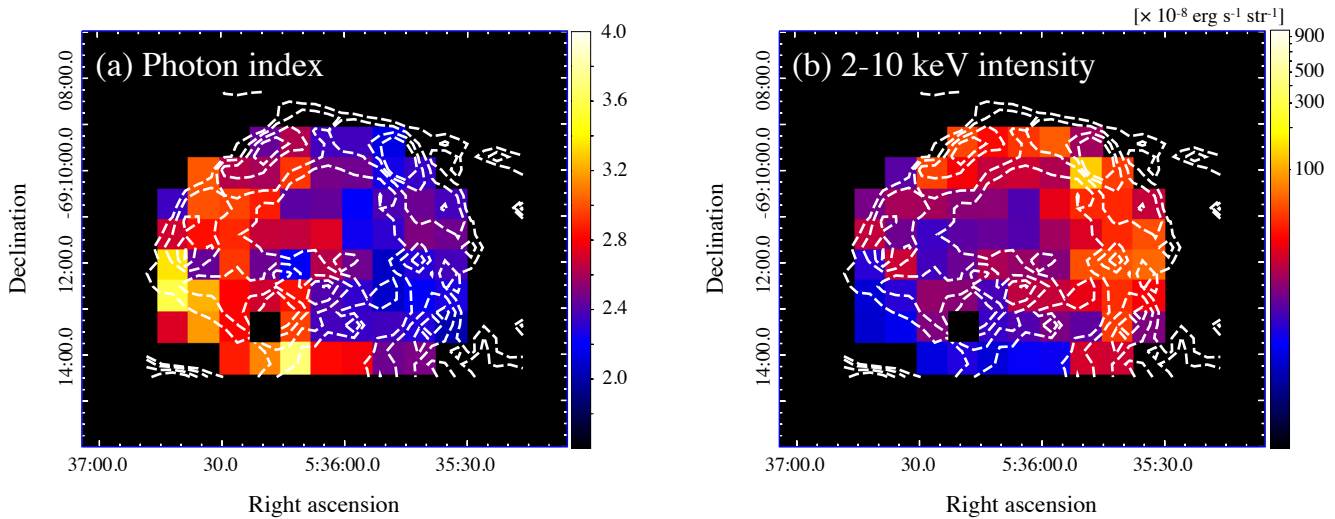}
  \end{center}
  \caption{Maps of the best-fit parameters: (a) the photon index $\Gamma$  and (b) absorption-corrected intensity in 2--10 keV [10$^{-8}$ erg s$^{-1}$ cm$^{-2}$ str$^{-1}$], of the non-thermal component. 
Smoothed white contours of the $XMM-newton$ EPIC-pn images (0.5--7 keV) are overlaid in the maps. 
}
  \label{fig:map}
\end{figure*}

Figures \ref{fig:map} (a) and (b) show the distributions of the photon index and absorption corrected 2--10 keV intensity of the non-thermal component, respectively. 
It is found for the first time that the non-thermal component is detected significantly in all the 70 regions covering the entire region of 30 Dor C.
Their typical relative error is $\sim$8\%. 
The photon index shows spatial variation of $\sim$2.0--3.7. 
The areas with the relatively steep / flat photon indices are distributed in the east / west regions, respectively. 
Even though this sort of high spatial resolution spectral analysis had not been performed so far, the trend is consistent with the previous studies \cite[e.g.,][]{2015A&A...573A..73K}.
The intensity in 2--10 keV significantly varies by more than an order of magnitude ($\sim$4–-130 $\times$ 10$^{-8}$ erg s$^{-1}$ cm$^{-2}$ str$^{-1}$) in the field and is relatively large in the west region of the shell structure.
Their typical relative error is $\sim$15\%. 

The temperature and normalization of the thermal component mainly emitting oxygen lines vary from $\sim$0.2 to $\sim$0.3 keV and from $\sim$0.2 to $\sim$7 [$10^{17}$ cm$^{-5}$ str$^{-1}$], respectively.
Because the low-temperature thermal plasma is not detected in the source-free region of the vicinity of 30 Dor C and the morphology apparently forms a shell-like structure as shown in figure \ref{fig:30_Dor_C_img}, the plasma may be associated with 30 Dor C.
Such low-temperature plasma with a temperature of $\sim$0.1--0.3 keV is found also in other SBs \cite[e.g.,][]{2001ApJS..136..119D,2010ApJ...715..412Y} and detected mainly in the east region as previously reported in  
\citet{2004ApJ...602..257B,2009PASJ...61S.175Y,2015A&A...573A..73K}. 
The temperature and normalization of the thermal component mainly emitting Fe L-shell lines vary from $\sim$0.9 to $\sim$1.2 keV and from $\sim$0.1 to $\sim$0.2 [$10^{17}$ cm$^{-5}$ str$^{-1}$], respectively. 
The temperature is consistent with that of the observed in the source-free region within the statistical error. 
According to the results of \citet{2002A&A...392..103S}, 
the flux of ISM in LMC varies by more than twice depending on the regions.
The normalization of the high-temperature components is consistent with that of the observed in background spectra within the variation.
While the results suggest that the high-temperature plasma may be due to the spatial variation of the ISM in the LMC,
the origin of the component is beyond our scope.
We confirmed that the uncertainty, e.g., in the metal abundance, does not affect our results for the non-thermal component significantly.

The intrinsic absorbing column density $N_{\rm H}$ of the LMC ranges from $\sim$0.3 to $\sim$2 $\times$ 10$^{22}$ cm$^{-2}$ and a typical relative error is $\sim$15 \%.
The intrinsic absorption in the east area of 30 Dor C seems to be relatively small ($\sim$0.6 $\times$ 10$^{22}$ cm$^{-2}$) while large ($\sim$1 $\times$ 10$^{22}$ cm$^{-2}$) in the west of the shell-like structure.

We confirmed that our representative results for our spectral analysis on the temperature of the thermal components, photon index of the non-thermal components, and intrinsic absorbing column density in the LMC are consistent with those of the previous studies \cite[e.g.,][]{2004ApJ...602..257B,2004ApJ...611..881S,2009PASJ...61S.175Y,2015A&A...573A..73K}. 

\input{bestfit_180709.tex}

\section{DISCUSSION}
We conducted the spatially resolved spectral analysis of 30 Dor C with a physical scale of $\sim$10 pc in X-ray for the first time. We revealed that the non-thermal emission exists in all the regions covering the whole area of 30 Dor C and extracted the distribution of the physical properties such as the photon index and absorption-corrected intensity of the non-thermal component. 
We found that the spectral shape changes and therefore the physical properties vary in this field.
In this section, we discussed, in particular, the origin of the spatial variation of the non-thermal X-ray properties to study the mechanism of cosmic-ray acceleration in SBs.

\begin{figure*}
  \begin{center}
    \includegraphics[width=170mm,angle=0]{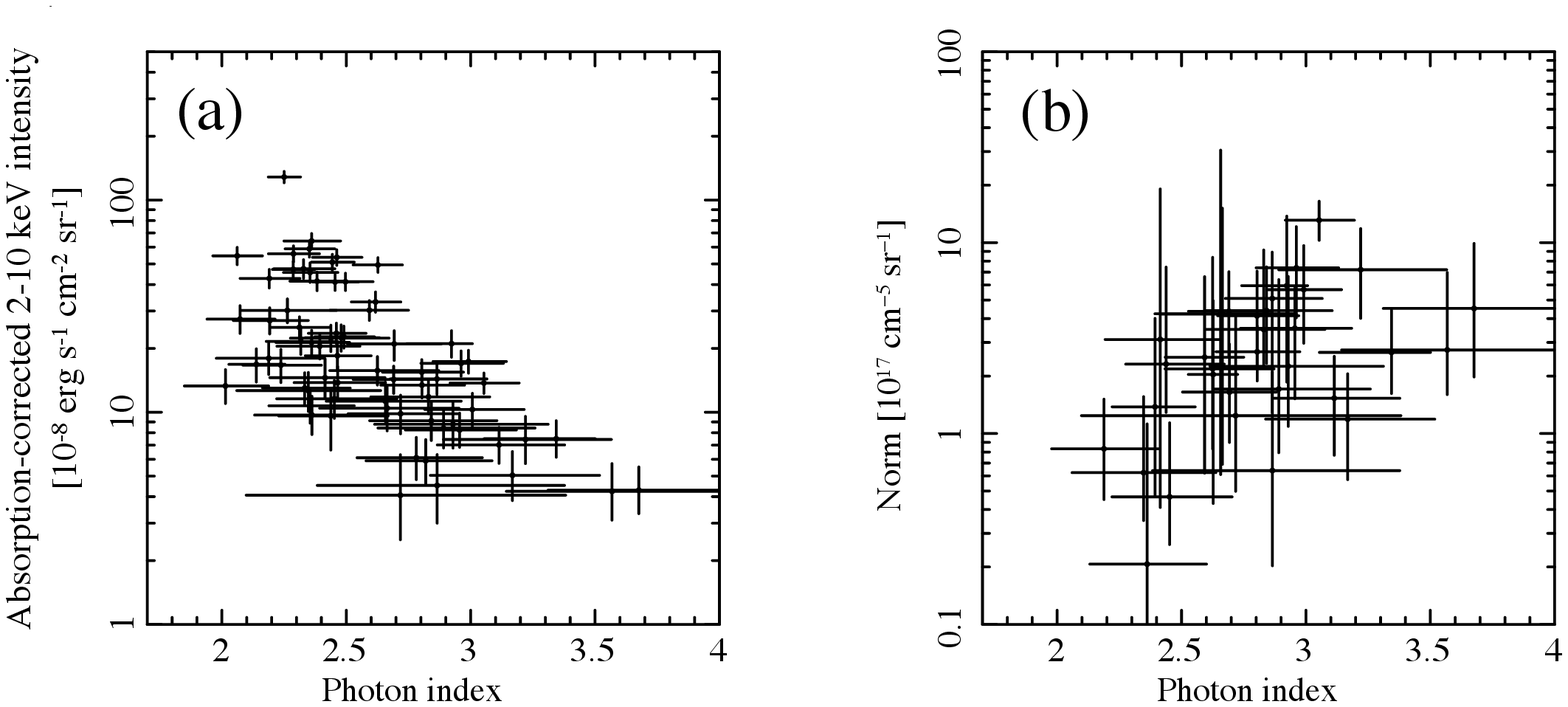}
  \end{center}
\vspace{-0cm}
\caption{Correlation plots for (a) the photon index vs. the 2--10 keV intensity [10$^{-8}$ erg s$^{-1}$ cm$^{-2}$], (b) Normalization of the low temperature thermal components [$10^{17}$ cm$^{-5}$ str$^{-1}$] vs. photon index, respectively.
 }

\label{fig:properties}
\end{figure*}

Some SNRs also show spatial variation of the photon index and intensity of the non-thermal component \citep[e.g.,][]{2015ApJ...799..175S,2017ApJ...835...34T}.
In particular, pc-scale spatially resolved spectral analysis reveals that the photon index closely correlates with the synchrotron X-ray intensity \citep[e.g.,][]{2015ApJ...799..175S}.
Thus, we also extracted the relation between the photon index and the synchrotron X-ray intensity in the same manner as shown in figure \ref{fig:properties}(a). 
There is a clear negative correlation with a correlation coefficient of $\sim$-0.5.
One of the interpretations is due to a shock-cloud interaction.
Magnetohydrodynamic numerical simulations in \cite{2012ApJ...744...71I} predict that such spatial variation can be produced by the shock-cloud interaction because the synchrotron X-ray intensity is positively correlated with the strength of the enhanced magnetic field due to the turbulence occurred in the interaction. 
The photon index also can be changed by the enhanced magnetic field since the particle acceleration occurs efficiently.
Therefore, it is naturally expected that the photon index gets flatter with increasing the synchrotron X-ray intensity.

According to the scenario, the larger the photon index is, the smaller the cut-off energy in the energy distribution of
electrons should be.  When we applied a broken power-law model instead of the power-law model for the non-thermal
X-ray emission, however, no constraint was given to the breaking energy with the X-ray spectra alone. The correlation
between the photon index and the cut-off energy has been observed in some SNRs when X-ray synchrotron spectra
are analyzed using the SRCUT model \citep{1998ApJ...493..375R,1999ApJ...525..368R} combined with radio synchrotron spectra
\citep{2004A&A...425..121R,2005ApJ...632..294B,2005ApJ...621..793B}. The spatially-resolved flux and spectral index
of the radio synchrotron emission of 30 Dor C have been obtained in \citet{2015A&A...573A..73K}, but as the authors say 
it is difficult to obtain their reliable values for the entire region of 30 Dor C due to the contaminations of 
a foreground molecular cloud and of thermal radio emission. This situation prevents us from analyzing the
multi-wavelength spectra from radio to X-ray.  The analysis of the spatially-resolved spectral energy distribution is
left to future works.

\cite{2017ApJ...843...61S} presents the molecular cloud distribution around 30 Dor C and it seems that there is a positive correlation between the synchrotron X-ray intensity and the amount of the molecular cloud.
The detailed comparison between X-ray and radio observations will be discussed (Yamane et al. in prep.).

\cite{2017ApJ...835...34T} argues that efficient acceleration occurs in the low density environment implying that the photon index steepens with increasing the normalization of the thermal component observationally based on the pc-scale spectral analysis results.
Thus, we also extracted the relation as shown in figure \ref{fig:properties}(b).
One can see a positive correlation with a correlation coefficient of $\sim$0.4 and thus similarities for SNRs are found in terms of the correlations between the non-thermal properties themselves and the non-thermal and thermal properties, which suggests the possibility that the same acceleration mechanism works also in the supperbubble.

\begin{figure}[t]
\vspace*{0.7cm}
  \begin{center}
    \includegraphics[width=90mm,angle=0]{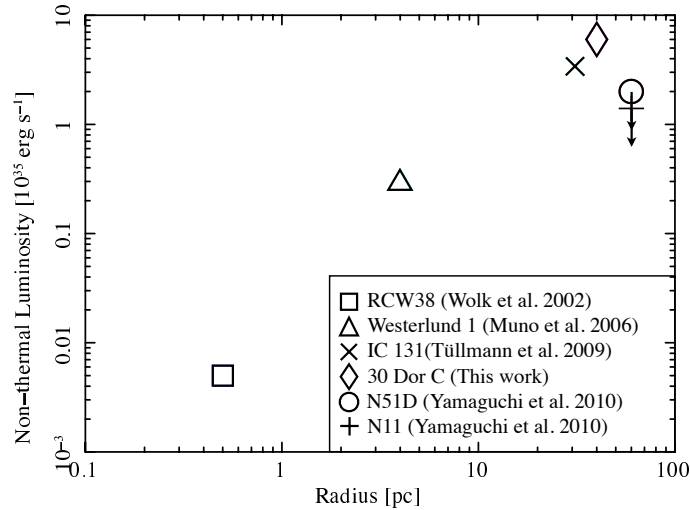}
  \end{center}
  \caption{
Non-thermal X-ray luminosity in 2--10 keV as a function of the radius for superbubbles. 
 }
  \label{fig:SB_nonthermal}
\end{figure}

The bright non-thermal X-ray emission in 30 Dor C was detected. 
However, no other SBs exhibit such a bright non-thermal emission. 
The SBs, where non-thermal X-ray emission has been significantly detected, 
are only RCW38 \citep{2002ApJ...580L.161W}, Westerlund 1 \citep{2006ApJ...650..203M} and IC 131 \citep{2009ApJ...707.1361T}. 
This sort of variation is observed also in SNRs and \cite{2012ApJ...746..134N} discussed the time evolution of the non-thermal component as a function of the radius which can be an indicator of the dynamical age of the SNR as described in \cite{1977ApJ...218..377W}.
As an analogy of the SNR case, we also investigated the relation between the non-thermal luminosity and the radius of the SBs as shown in figure \ref{fig:SB_nonthermal}. 
The non-thermal luminosity goes up with increasing the radius up to $\sim$40 pc, whereas it then appears to be decreases. 
30 Dor C is located around the peak, which suggests that the system is currently on a phase of high energy particle acceleration.

\section{SUMMARY}
We conducted a detailed spatial analysis using the large amount of $XMM-Newton$ archival data for 30 Dor C to study spatial variation of mainly the non-thermal component.
The 30 Dor C field was divided into 70 regions with a physical scale of $\sim$10 pc and we found for the first time that the non-thermal emission exists in all the regions covering the whole field of 30 Dor C.
The extracted spectra in the east region can be described well with
a combination of the thermal and non-thermal models,
whereas the spectra in the west region can be well fitted with
the non-thermal model alone. 
The photon index and intensity in 2--10 keV indicate the spatial variation of $\sim$2.0--3.7 and $\sim$(4--130) $\times$ 10$^{-8}$ erg s$^{-1}$ cm$^{-2}$ str$^{-1}$) in the field and the negative correlation between the non-thermal physical properties is observed.
The temperature and normalization of the thermal component also vary within a range of $\sim$0.2--0.3 keV 
and $\sim$0.2--7 $\times$ 10$^{17}$ cm$^{-5}$ str$^{-1}$, respectively. 
The positive correlation between the photon index and the normalization of the thermal component is also observed as is the case in SNRs, suggesting that the same acceleration mechanism dominates also in the supperbubble.\\

This research was supported by a grant from the Hayakawa Satio Fund awarded by the Astronomical Society of Japan.
HM is supported by JSPS KAKENHI Grand Number JP 15640356.
IM acknowledge supports from the by JSPS KAKENHI Grand Number JP 26220703.
HS is supported by JSPS KAKENHI Grand Number JP 16K17664.
The authors are grateful to the anonymous referee for his/her comprehensive comments and useful suggestions, which improved the paper very much.

\end{document}

%% file: bestfit_180709.tex
\begin{deluxetable*}{ccccccccc}[b!]
\tablecaption{Best fit parameters of fitting \label{tab:fitting_results}.}
\tablecolumns{9}
\tablenum{3}
\tablewidth{0pt}
\tablehead{
\colhead{} &
\colhead{LMC absorption } &
\multicolumn{2}{c}{Power-law component} & \multicolumn{2}{c}{Low temperature component} & \multicolumn{2}{c}{High temperature component} & \colhead{}\\
\colhead{Region num.} & \colhead{$N_{\rm H, LMC}$\tablenotemark{a}} & \colhead{$\Gamma$} & \colhead{2-10 keV intensity\tablenotemark{b}} & 
\colhead{$kT$\tablenotemark{c}} & \colhead{$Norm$\tablenotemark{d}} & \colhead{$kT$\tablenotemark{c}} & \colhead{$Norm$\tablenotemark{d}} & $\chi^2/d.o.f$  
}
\startdata
1 &0.26$_{-0.12}^{+0.20}$ &2.35$_{-0.29}^{+0.29}$ &12.6$_{-2.6}^{+2.8}$ &0.26$_{-0.04}^{+0.03}$ &0.62$_{-0.27}^{+0.94}$ &1.09$_{-0.06}^{+0.18}$ &0.12$_{-0.05}^{+0.06}$ &244/211\\ 
2 &0.37$_{-0.11}^{+0.12}$ &2.69$_{-0.19}^{+0.19}$ &21.0$_{-3.5}^{+3.2}$ &0.25$_{-0.03}^{+0.03}$ &1.65$_{-0.75}^{+1.3}$ &1.19$_{-0.25}^{+0.41}$ &0.14$_{-0.08}^{+0.10}$ &253/300\\ 
3 &0.47$_{-0.11}^{+0.10}$ &3.34$_{-0.29}^{+0.16}$ &7.5$_{-1.4}^{+1.6}$ &0.23$_{-0.02}^{+0.03}$ &2.7$_{-1.0}^{+1.8}$ &--&--&255/268\\ 
4 &0.54$_{-0.12}^{+0.18}$ &3.57$_{-0.42}^{+0.43}$ &4.2$_{-1.1}^{+1.5}$ &0.23$_{-0.04}^{+0.03}$ &2.7$_{-1.1}^{+4.2}$ &--&--&160/160\\ 
5 &0.56$_{-0.22}^{+0.27}$ &2.72$_{-0.62}^{+0.66}$ &4.1$_{-1.6}^{+2.2}$ &0.24$_{-0.05}^{+0.05}$ &1.24$_{-0.74}^{+2.9}$ &--&--&98/127\\ 
6 &0.33$_{-0.08}^{+0.08}$ &3.01$_{-0.19}^{+0.21}$ &10.3$_{-1.8}^{+2.0}$ &--&--&--&--&174/162\\ 
7 &0.68$_{-0.11}^{+0.10}$ &2.99$_{-0.14}^{+0.15}$ &17.3$_{-2.1}^{+2.2}$ &0.19$_{-0.01}^{+0.02}$ &5.7$_{-2.7}^{+4.0}$ &--&--&271/274\\ 
8 &0.60$_{-0.11}^{+0.20}$ &2.83$_{-0.23}^{+0.25}$ &11.8$_{-2.0}^{+2.2}$ &0.23$_{-0.04}^{+0.02}$ &3.5$_{-1.3}^{+5.6}$ &--&--&254/249\\ 
9 &0.50$_{-0.13}^{+0.23}$ &2.44$_{-0.16}^{+0.23}$ &22.4$_{-3.0}^{+3.4}$ &0.23$_{-0.04}^{+0.02}$ &2.3$_{-1.0}^{+5.1}$ &--&--&281/235\\ 
10 &0.63$_{-0.12}^{+0.11}$ &3.22$_{-0.33}^{+0.35}$ &7.4$_{-1.7}^{+2.1}$ &0.19$_{-0.01}^{+0.02}$ &7.2$_{-3.2}^{+4.6}$ &--&--&185/177\\ 
11 &0.35$_{-0.15}^{+0.12}$ &3.17$_{-0.33}^{+0.35}$ &5.0$_{-1.2}^{+1.5}$ &0.25$_{-0.02}^{+0.05}$ &1.19$_{-0.62}^{+0.86}$ &--&--&183/193\\ 
12 &0.58$_{-0.13}^{+0.13}$ &2.63$_{-0.10}^{+0.10}$ &49.5$_{-3.9}^{+4.1}$ &0.19$_{-0.02}^{+0.05}$ &2.0$_{-1.6}^{+2.9}$ &--&--&380/369\\ 
13 &0.74$_{-0.11}^{+0.10}$ &2.96$_{-0.16}^{+0.17}$ &17.0$_{-2.2}^{+2.5}$ &0.19$_{-0.01}^{+0.02}$ &7.4$_{-3.3}^{+4.8}$ &--&--&239/251\\ 
14 &0.50$_{-0.16}^{+0.26}$ &2.89$_{-0.26}^{+0.37}$ &8.4$_{-1.6}^{+1.9}$ &0.23$_{-0.05}^{+0.04}$ &1.71$_{-0.92}^{+4.7}$ &--&--&198/185\\ 
15 &0.53$_{-0.16}^{+0.22}$ &2.93$_{-0.31}^{+0.38}$ &8.8$_{-2.0}^{+2.4}$ &0.23$_{-0.04}^{+0.03}$ &2.3$_{-1.2}^{+4.4}$ &--&--&172/172\\ 
16 &0.53$_{-0.13}^{+0.10}$ &2.80$_{-0.16}^{+0.17}$ &15.5$_{-2.1}^{+2.2}$ &0.21$_{-0.02}^{+0.03}$ &4.1$_{-1.9}^{+3.0}$ &--&--&271/249\\ 
17 &0.41$_{-0.08}^{+0.17}$ &2.80$_{-0.17}^{+0.17}$ &13.4$_{-1.7}^{+1.9}$ &0.23$_{-0.01}^{+0.02}$ &2.68$_{-0.79}^{+0.98}$ &--&--&359/315\\ 
18 &0.29$_{-0.20}^{+0.24}$ &2.87$_{-0.48}^{+0.51}$ &4.5$_{-1.5}^{+1.8}$ &0.23$_{-0.04}^{+0.06}$ &0.64$_{-0.44}^{+1.5}$ &0.93$_{-0.15}^{+0.19}$ &0.11$_{-0.04}^{+0.04}$ &163/156\\ 
19 &0.35$_{-0.04}^{+0.05}$ &2.44$_{-0.08}^{+0.09}$ &50.9$_{-4.3}^{+4.6}$ &--&--&--&--&293/291\\ 
20 &0.51$_{-0.19}^{+0.19}$ &2.59$_{-0.16}^{+0.16}$ &30.3$_{-3.2}^{+3.4}$ &0.19$_{-0.02}^{+0.03}$ &2.5$_{-1.9}^{+4.1}$ &0.95$_{-0.15}^{+0.26}$ &0.22$_{-0.08}^{+0.07}$ &321/317\\ 
21 &0.80$_{-0.25}^{+0.14}$ &2.86$_{-0.19}^{+0.20}$ &14.4$_{-2.2}^{+2.5}$ &0.19$_{-0.01}^{+0.05}$ &5.1$_{-3.8}^{+3.8}$ &--&--&222/214\\ 
22 &0.94$_{-0.36}^{+0.20}$ &2.66$_{-0.27}^{+0.29}$ &10.4$_{-2.3}^{+2.7}$ &0.19$_{-0.04}^{+0.08}$ &4.2$_{-3.4}^{+11}$ &--&--&133/125\\ 
23 &0.24$_{-0.10}^{+0.19}$ &2.45$_{-0.23}^{+0.25}$ &11.6$_{-2.2}^{+2.6}$ &0.30$_{-0.06}^{+0.05}$ &0.47$_{-0.20}^{+0.67}$ &--&--&209/187\\ 
24 &0.32$_{-0.11}^{+0.10}$ &2.69$_{-0.16}^{+0.17}$ &14.3$_{-2.0}^{+2.2}$ &0.19$_{-0.01}^{+0.01}$ &4.4$_{-1.9}^{+2.7}$ &--&--&289/277\\ 
25\tablenotemark{e} & --&-- &-- &--&--&-- &-- &--\\ 
26 &0.44$_{-0.14}^{+0.11}$ &3.11$_{-0.25}^{+0.26}$ &7.0$_{-1.3}^{+1.5}$ &0.24$_{-0.02}^{+0.04}$ &1.53$_{-0.76}^{+1.0}$ &--&--&233/210\\ 
27 &0.42$_{-0.05}^{+0.06}$ &2.62$_{-0.10}^{+0.10}$ &33.1$_{-3.5}^{+3.8}$ &--&--&--&--&260/258\\ 
28 &0.78$_{-0.19}^{+0.14}$ &2.92$_{-0.18}^{+0.08}$ &21.2$_{-2.9}^{+3.2}$ &0.17$_{-0.02}^{+0.03}$ &6.0$_{-4.1}^{+7.8}$ &--&--&212/207\\ 
29 &0.82$_{-0.39}^{+0.28}$ &2.41$_{-0.22}^{+0.24}$ &14.6$_{-2.9}^{+3.3}$ &0.19$_{-0.04}^{+0.07}$ &3.1$_{-2.7}^{+16}$ &--&--&112/128\\ 
30 &0.91$_{-0.37}^{+0.30}$ &2.66$_{-0.24}^{+0.30}$ &11.3$_{-2.2}^{+2.6}$ &0.18$_{-0.05}^{+0.07}$ &4.2$_{-3.6}^{+26}$ &--&--&149/146\\ 
31 &0.32$_{-0.14}^{+0.15}$ &2.19$_{-0.21}^{+0.22}$ &18.0$_{-2.9}^{+3.2}$ &0.30$_{-0.04}^{+0.05}$ &0.83$_{-0.38}^{+0.68}$ &--&--&153/188\\ 
32 &0.49$_{-0.12}^{+0.11}$ &2.84$_{-0.25}^{+0.26}$ &9.1$_{-1.8}^{+2.1}$ &0.19$_{-0.01}^{+0.02}$ &4.4$_{-2.1}^{+3.1}$ &--&--&250/207\\ 
33 &0.47$_{-0.18}^{+0.10}$ &2.96$_{-0.22}^{+0.23}$ &8.3$_{-1.4}^{+1.7}$ &0.20$_{-0.01}^{+0.04}$ &3.6$_{-2.0}^{+2.3}$ &--&--&245/249\\ 
34 &0.73$_{-0.14}^{+0.13}$ &3.68$_{-0.37}^{+0.37}$ &4.3$_{-1.0}^{+1.2}$ &0.18$_{-0.02}^{+0.02}$ &4.5$_{-2.5}^{+5.4}$ &--&--&211/166\\ 
35 &0.49$_{-0.07}^{+0.07}$ &2.35$_{-0.11}^{+0.11}$ &45.6$_{-4.7}^{+5.0}$ &--&--&--&--&193/209\\ 
36 &0.52$_{-0.08}^{+0.09}$ &2.48$_{-0.14}^{+0.15}$ &22.5$_{-3.1}^{+3.3}$ &--&--&--&--&177/152\\ 
37 &0.57$_{-0.13}^{+0.15}$ &2.44$_{-0.21}^{+0.23}$ &9.6$_{-3.0}^{+3.4}$ &--&--&--&--&126/117\\ 
38 &0.58$_{-0.12}^{+0.13}$ &2.72$_{-0.21}^{+0.23}$ &9.8$_{-1.9}^{+2.2}$ &--&--&--&--&145/138\\ 
39 &0.82$_{-0.21}^{+0.26}$ &2.63$_{-0.19}^{+0.25}$ &15.7$_{-2.3}^{+2.6}$ &0.24$_{-0.05}^{+0.05}$ &2.2$_{-1.4}^{+6.2}$ &--&--&185/186\\ 
40 &0.46$_{-0.19}^{+0.20}$ &2.39$_{-0.17}^{+0.16}$ &20.5$_{-2.7}^{+2.9}$ &0.21$_{-0.03}^{+0.04}$ &1.4$_{-0.9}^{+2.6}$ &--&--&170/187\\ 
\enddata
\end{deluxetable*}

\begin{deluxetable*}{ccccccccc}[t!]
\tablecaption{Best fit parameters of fitting \label{tab:fitting_results}.}
\tablecolumns{9}
\tablenum{3}
\tablewidth{0pt}
\tablehead{
\colhead{} &
\colhead{LMC absorption} &
\multicolumn{2}{c}{Power-law component} & \multicolumn{2}{c}{Low temperature component} & \multicolumn{2}{c}{High temperature component} & \colhead{}\\
\colhead{Region num.} & \colhead{$N_{\rm H, LMC}$\tablenotemark{a}} & \colhead{$\Gamma$} & \colhead{2-10 keV intensity\tablenotemark{b}} & 
\colhead{$kT$\tablenotemark{c}} & \colhead{$Norm$\tablenotemark{d}} & \colhead{$kT$\tablenotemark{c}} & \colhead{$Norm$\tablenotemark{d}} & $\chi^2/d.o.f$  
}
\startdata
41 &0.34$_{-0.15}^{+0.35}$ &2.36$_{-0.23}^{+0.24}$ &9.7$_{-1.9}^{+2.2}$ &0.33$_{-0.11}^{+0.16}$ &0.21$_{-0.14}^{+0.92}$ &--&--&184/156\\ 
42 &0.33$_{-0.10}^{+0.11}$ &2.82$_{-0.24}^{+0.27}$ &5.9$_{-1.3}^{+1.5}$ &--&--&--&--&115/113\\ 
43 &0.80$_{-0.08}^{+0.08}$ &2.35$_{-0.10}^{+0.10}$ &58.9$_{-5.1}^{+5.3}$ &--&--&--&--&231/242\\ 
44 &0.79$_{-0.10}^{+0.11}$ &2.46$_{-0.13}^{+0.14}$ &18.5$_{-2.8}^{+3.0}$ &--&--&--&--&151/162\\ 
45 &0.64$_{-0.12}^{+0.13}$ &2.19$_{-0.15}^{+0.16}$ &27.1$_{-3.7}^{+4.0}$ &--&--&--&--&142/150\\ 
46 &0.55$_{-0.11}^{+0.12}$ &2.24$_{-0.15}^{+0.16}$ &16.7$_{-2.9}^{+3.1}$ &--&--&--&--&150/145\\ 
47 &0.52$_{-0.10}^{+0.11}$ &2.47$_{-0.18}^{+0.19}$ &13.8$_{-2.4}^{+2.6}$ &--&--&--&--&143/146\\ 
48 &0.44$_{-0.08}^{+0.09}$ &2.32$_{-0.14}^{+0.15}$ &21.6$_{-2.8}^{+3.1}$ &--&--&--&--&142/163\\ 
49 &0.43$_{-0.10}^{+0.11}$ &2.33$_{-0.17}^{+0.18}$ &13.0$_{-2.1}^{+2.4}$ &--&--&--&--&167/143\\ 
50 &0.49$_{-0.12}^{+0.13}$ &2.78$_{-0.24}^{+0.27}$ &6.1$_{-1.3}^{+1.5}$ &--&--&--&--&105/98\\ 
51 &0.84$_{-0.10}^{+0.11}$ &2.14$_{-0.11}^{+0.12}$ &16.8$_{-3.0}^{+3.1}$ &--&--&--&--&200/199\\ 
52 &0.83$_{-0.06}^{+0.06}$ &2.25$_{-0.06}^{+0.07}$ &128.4$_{-7.8}^{+8.0}$ &--&--&--&--&387/393\\ 
53 &0.90$_{-0.09}^{+0.10}$ &2.38$_{-0.11}^{+0.11}$ &41.5$_{-4.1}^{+4.3}$ &--&--&--&--&207/223\\ 
54 &0.92$_{-0.11}^{+0.12}$ &2.31$_{-0.12}^{+0.12}$ &25.1$_{-2.9}^{+3.1}$ &--&--&--&--&210/191\\ 
55 &0.99$_{-0.11}^{+0.12}$ &2.06$_{-0.10}^{+0.10}$ &54.6$_{-5.0}^{+5.2}$ &--&--&--&--&229/226\\ 
56 &0.49$_{-0.10}^{+0.10}$ &2.07$_{-0.13}^{+0.14}$ &27.5$_{-4.0}^{+4.3}$ &--&--&--&--&162/155\\ 
57 &0.54$_{-0.11}^{+0.12}$ &2.35$_{-0.17}^{+0.18}$ &10.7$_{-1.8}^{+2.0}$ &--&--&--&--&149/142\\ 
58 &0.40$_{-0.06}^{+0.06}$ &2.46$_{-0.11}^{+0.12}$ &23.6$_{-2.6}^{+2.7}$ &--&--&--&--&196/205\\ 
59 &1.10$_{-0.12}^{+0.13}$ &2.33$_{-0.12}^{+0.13}$ &47.4$_{-4.7}^{+5.0}$ &--&--&--&--&178/183\\ 
60 &0.98$_{-0.08}^{+0.08}$ &2.46$_{-0.09}^{+0.09}$ &41.1$_{-3.5}^{+3.7}$ &--&--&--&--&272/284\\ 
61 &1.14$_{-0.10}^{+0.11}$ &2.50$_{-0.11}^{+0.11}$ &41.3$_{-3.8}^{+4.0}$ &--&--&--&--&284/249\\ 
62 &1.55$_{-0.14}^{+0.15}$ &2.29$_{-0.10}^{+0.10}$ &55.8$_{-4.8}^{+4.9}$ &--&--&--&--&286/259\\ 
63 &0.95$_{-0.12}^{+0.13}$ &2.19$_{-0.12}^{+0.12}$ &42.7$_{-4.2}^{+4.4}$ &--&--&--&--&206/191\\ 
64 &0.54$_{-0.06}^{+0.06}$ &2.29$_{-0.08}^{+0.09}$ &47.1$_{-4.1}^{+4.3}$ &--&--&--&--&278/266\\ 
65 &0.32$_{-0.06}^{+0.06}$ &2.49$_{-0.12}^{+0.12}$ &22.7$_{-2.6}^{+2.8}$ &--&--&--&--&194/198\\ 
66 &0.84$_{-0.13}^{+0.14}$ &2.36$_{-0.14}^{+0.15}$ &21.3$_{-2.8}^{+3.0}$ &--&--&--&--&166/163\\ 
67 &1.36$_{-0.11}^{+0.12}$ &2.46$_{-0.10}^{+0.10}$ &53.7$_{-4.4}^{+4.6}$ &--&--&--&--&295/274\\ 
68 &1.99$_{-0.18}^{+0.20}$ &2.36$_{-0.11}^{+0.12}$ &64.1$_{-5.2}^{+5.4}$ &--&--&--&--&236/248\\ 
69 &1.69$_{-0.28}^{+0.32}$ &2.26$_{-0.18}^{+0.20}$ &30.2$_{-3.8}^{+4.0}$ &--&--&--&--&127/137\\ 
70 &0.33$_{-0.10}^{+0.11}$ &2.01$_{-0.17}^{+0.17}$ &13.3$_{-2.3}^{+2.6}$ &--&--&--&--&261/232\\ 
\enddata
\tablenotetext{a}{The unit is 10$^{22}$ cm$^{-2}$.}
\tablenotetext{b}{The unit is 10$^{-8}$ erg s$^{-1}$ cm$^{-2}$ sr$^{-1}$.}
\tablenotetext{c}{The unit is keV.}
\tablenotetext{d}{The unit is 10$^{17}$ cm$^{-5}$ sr$^{-1}$.}
\tablenotetext{e}{This region is excluded in our analysis and discussion due to a heavy contamination from a young SNR MCSNR~J0536-6913 described well with a combination of non-thermal and NEI models.}
\end{deluxetable*}